\documentclass[journal=acsnano,manuscript=article]{achemso}

\usepackage[version=3]{mhchem}
\usepackage{xcolor}
\usepackage{ulem}

\author{Jennifer Schmeink}
\affiliation[University of Duisburg-Essen]
{Faculty of Physics and CENIDE, University of Duisburg-Essen, Duisburg, Germany}
\author{Vladislav Musytschuk}
\affiliation[University of Duisburg-Essen]
{Faculty of Physics and CENIDE, University of Duisburg-Essen, Duisburg, Germany}
\author{Erik Pollmann}
\affiliation[University of Duisburg-Essen]
{Faculty of Physics and CENIDE, University of Duisburg-Essen, Duisburg, Germany}
\author{Stephan Sleziona}
\affiliation[University of Duisburg-Essen]
{Faculty of Physics and CENIDE, University of Duisburg-Essen, Duisburg, Germany}
\author{André Maas}
\affiliation[University of Duisburg-Essen]
{Faculty of Physics and CENIDE, University of Duisburg-Essen, Duisburg, Germany}
\author{Marika Schleberger}
\email{marika.schleberger@uni-due.de}
\phone{+49 (0)203 379-1600}
\affiliation[University of Duisburg-Essen]
{Faculty of Physics and CENIDE, University of Duisburg-Essen, Duisburg, Germany}

\title[An \textsf{achemso} demo]
  {Lifetime of Excitons in Janus Monolayer MoSSe Prepared from Exfoliated MoSe$_2$}

\abbreviations{2D,TMDC,CVD,PL,AFM,RT,TCSPC}


\begin{document}

%%%%%%%%%%%%%%%%%%%%%%%%%TableOfContent%%%%%%%%%%%%%%%%%%%%%%%%%%%%%%
%\begin{tocentry}
%\includegraphics[width=\textwidth]{TOC_graphic.png}
%\end{tocentry}

%%%%%%%%%%%%%%%%%%%%%%%%%%%%Abstract%%%%%%%%%%%%%%%%%%%%%%%%%%%%%%%%%
\begin{abstract}
Janus monolayer transition metal dichalcogenides, where one of the two chalcogen layers is substituted with a different kind of chalcogen atoms, are pushing the properties of two dimensional materials into new territories. Yet only little is known about this new kind of material class, mainly due to the difficult synthesis. In this work we propose a method to prepare high quality Janus MoSSe monolayers from as-exfoliated MoSe$_2$ by thermal sulfurization. With this we aim to pave a way for more exotic Janus monolayers, which have been out of the experimental reach thus far. The synthesized MoSSe is diligently characterized by room- and low-temperature Raman and photoluminescence spectroscopy, atomic force microscopy correlated with Raman mappings, and time-correlated single-photon counting. The latter providing new information on the lifetime of excitons in Janus MoSSe monolayers. In addition, we report an enhanced trion formation at low temperatures and a relatively high excitonic transition energy that is indicative of less defect states and strain, and therefore a high sample quality. 
\end{abstract}
\newpage
%%%%%%%%%%%%%%%%%%%%%%%%%%%MainText%%%%%%%%%%%%%%%%%%%%%%%%%%%%%%%%%%
\section{Introduction}
Since the discovery of graphene in 2004, the field of two dimensional (2D) materials has matured and virtually dozens of 2D materials have been synthesized and characterized. The search for more versatile 2D materials has recently led to the successful synthesis of so-called Janus-type 2D materials \cite{Lu.2017,Zhang.2017,Sant.2020} based on transition metal dichalcogenide (TMDC) monolayers. Those Janus-type group-VI dichalcogenides are comprised of a TMDC where the chalcogen atoms on one side of the monolayer are substituted by another type of chalcogen atoms, typically denoted as MXY (M = Mo, W, ...; X, Y = S, Se, Te, X $\neq$ Y), see Figs. \ref{fig:structure+method} (a) and (b).
The top and bottom layer are thus made up from elements with different electronegativity which leads to a built-in electrical field within the Janus layer so that the out-of-plane mirror symmetry of the conventional MX$_2$ TMDC is broken resulting in completely different physical properties \cite{Li.2018,Zhang.2020,Wei.2020}. These Janus-type group-VI TMDC monolayers are predicted to be semiconductors, to preserve spin splitting at band edges and thus valley properties, to possess a large momentum dependent spin splitting (Rashba effect) due to the spontaneous out-of-plane dipole and an intrinsic in-plane piezoelectric effect \cite{RiisJensen.2019,Dong.2017b,Ahammed.2020,Rawat.2020}. Furthermore, the structural asymmetry along the out-of-plane direction enables efficient charge separation rendering the Janus TMDC monolayer attractive for photovoltaics and photocatalysis alike \cite{Ji.2018,Din.2019,Idrees.2019}. Two-dimensional layers of this type were already predicted to be stable as early on as 2013 \cite{Cheng.2013} but the synthesis of a Janus MoSSe monolayer was experimentally realized only recently in 2017 \cite{Lu.2017,Zhang.2017} and the number of different Janus materials synthesized since then has remained strikingly low.

The synthesis protocols reported so far, ranging from different kinds of plasma etching \cite{Lu.2017,Li.2020,Trivedi.2020,Zhang.2020b,Petric.2021}, to simple thermal etching \cite{Zhang.2017} in a chalcogen rich atmosphere, have as a common element been using a base material that was grown via chemical vapor deposition (CVD). While the general idea to replace one chalcogen layer via postprocessing has proven fruitful, the restriction to CVD grown base materials severely limits the number of possible Janus monolayers. 

To overcome this problem, we have established a fabrication method akin to Zhang et al. \cite{Zhang.2017}, but instead of using CVD-grown monolayer samples, we use as-exfoliated monolayers. While CVD-grown samples are larger in size, using exfoliated monolayers offers several advantages. Exfoliated samples are better suited with respect to accessibility and quality, but most notably they introduce a much wider range of possible base materials, which will help to advance the field of Janus-type TMDCs even further as more exotic materials become available. The numerous van der Waals bulk materials which already exist or have been identified \cite{Duong.2017,Marzari.2018,Frisenda.2020} and should be readily mechanically exfoliable, constitute a great starting point for new Janus-type structures. Moreover, the quality of bulk materials often exceeds that of common CVD materials. The latter are usually prone to intrinsic defects \cite{Hong.2015} and they typically exhibit a much stronger interaction with the chosen substrate than exfoliated samples leading to strain \cite{Pollmann.2018,Pollmann.2020}. This in return affects the Janus material's properties, such as its band gap structure or its excitonic behaviour. In our paper we address the challenge to synthesize high quality Janus materials and present a successful synthesis route starting from high quality base material.

\section{Results and discussion}

\begin{figure}[ht]
    \centering
    \includegraphics[width=\textwidth]{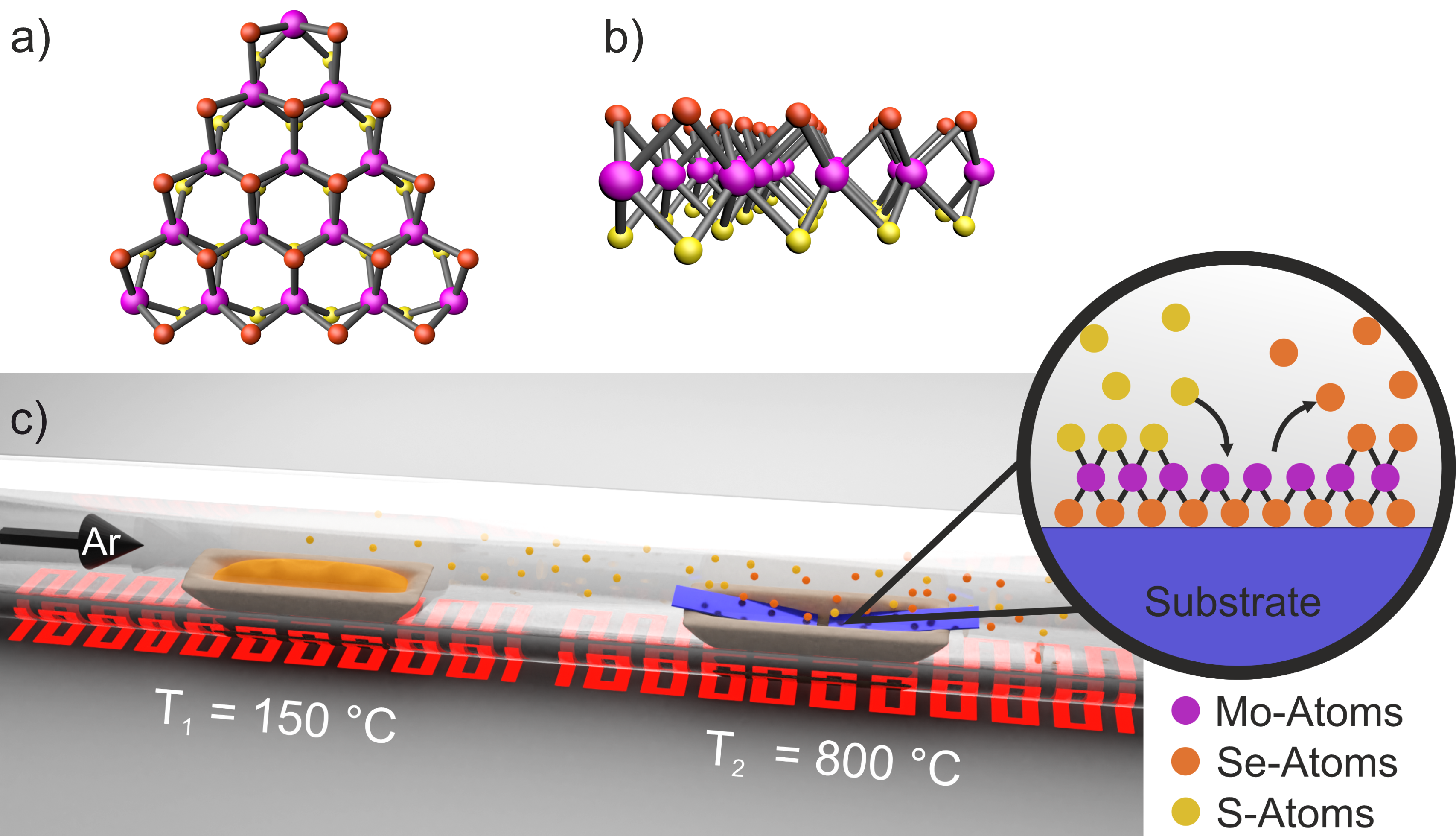}
    \caption{Top- (a) and side-view (b) of a Janus-type TMDC. Transition metal atoms (purple) are sandwiched between two layers of different chalcogen atoms (yellow \& orange). \linebreak (c) Schematic illustration of the thermal sulfurization process. Depicted are two heating zones of a tube furnace. The sulfur source and the exfoliated substrates are located in different, thermally isolated parts in the oven. Argon gas is inserted into the quartz tube coming from the left. The inset shows simplified the replacement of atoms during the thermal processing.}
    \label{fig:structure+method}
\end{figure}

We start by briefly describing our basic setup that is schematically shown in Fig. \ref{fig:structure+method} (c). We use a multi zone tube furnace to thermally etch exfoliated MoSe$_2$ samples in a sulfur rich environment. For more details see the method section. This approach works best with a selenium based material such as MoSe$_2$ because Se-bonds are weaker than S-bonds due to the longer bond length and the more delocalized electronic structure \cite{Chu.2016}. \newpage

\begin{figure}[htbp]
    \centering
    \includegraphics[width=0.9\textwidth]{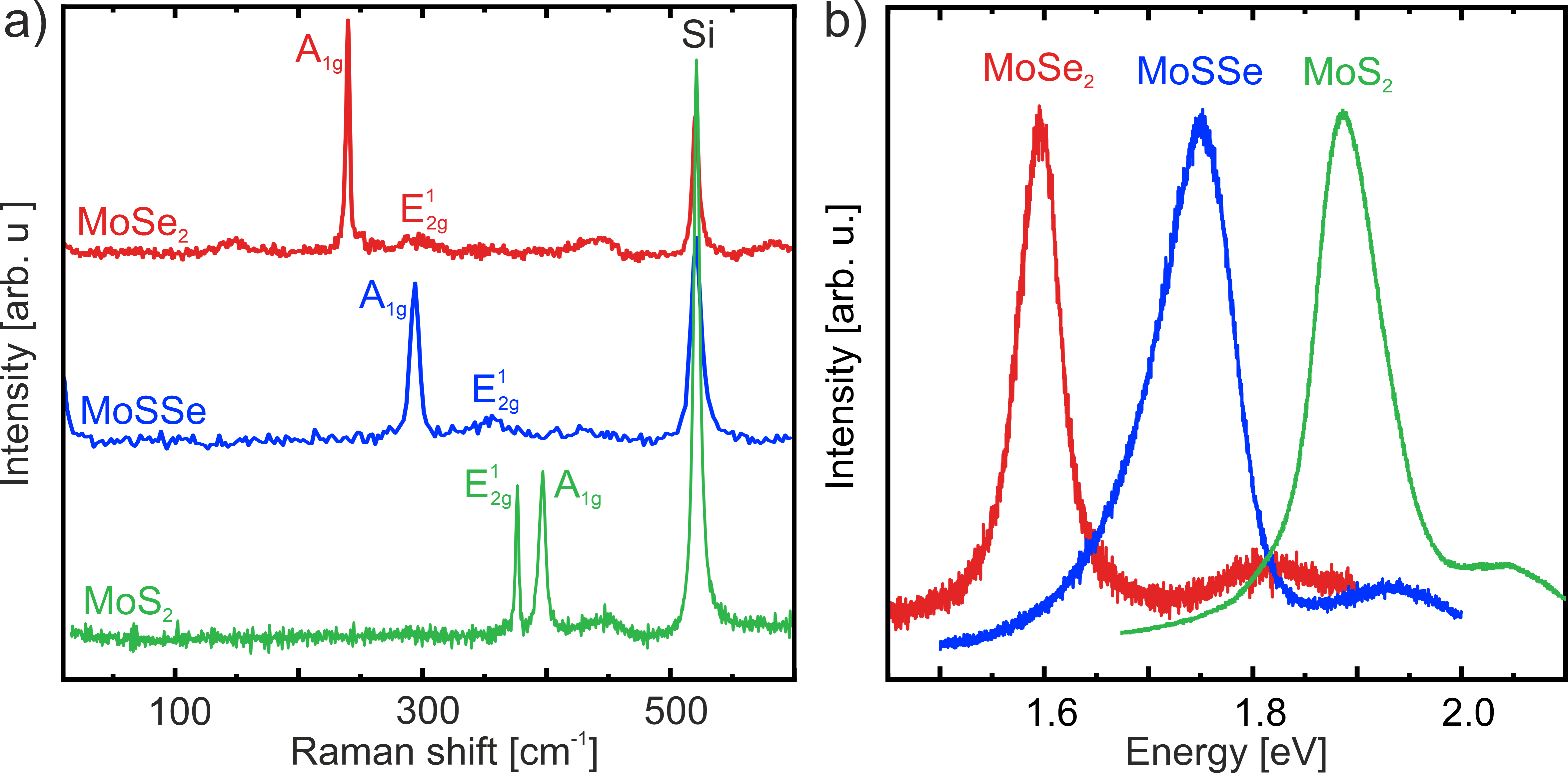}
    \caption{Successful preparation of a 2D Janus material starting from exfoliated MoSe$_2$. Normalized Raman- (a) and low temperature photoluminescence-spectra (b) comparison of the exfoliated base TMDCs: MoS$_2$ (green) and MoSe$_2$ (red), with the Janus monolayer MoSSe (blue). Raman spectra taken in the monolayer region show the characteristic A$_{1g}$ and E$^1_{2g}$ modes of MoSe$_2$. After thermal processing of an MoSe$_2$ sample in an S-rich atmosphere its Raman modes shift, indicating the successful transformation into Janus MoSSe.}
    \label{fig:Raman+PL}
\end{figure}

Before thermal processing, the monolayer regions of the exfoliated MoSe$_2$ samples are identified by atomic force microscopy as will be shown further down. The fastest method to ensure that the transformation process was successful is to inspect the samples' Raman and photoluminescence (PL) signatures. As shown in Fig. \ref{fig:Raman+PL} (a), the Raman signature of the Janus-Monolayer (blue) is distinctively different for the two base TMDCs MoS$_2$ (green) and MoSe$_2$ (red). All three materials show the typical in- and out-of-plane vibrational modes, denoted E$^1_{2g}$ and A$_{1g}$, respectively.
 As can be seen from the MoSSe spectrum, there are no remnants of the MoSe$_2$- nor MoS$_2$-modes. Therefore, the blue spectrum stems from a perfect Janus MoSSe monolayer with its A$_{1g}$-mode at 289~cm$^{-1}$ and its E$^1_{2g}$-mode at 352~cm$^{-1}$. These values are in good agreement with both, other experimental data \cite{Lu.2017,Zhang.2017} and theoretical results \cite{Xia.2018,Idrees.2019,Petric.2021}. The strong PL signals from MoSe$_2$ and MoS$_2$, see Fig.~\ref{fig:Raman+PL} (b), are by themselves a sign for the material being a monolayer, as they occur due to the shift from an indirect to a direct band gap when the materials are thinned down. Moreover, the energies of the two PL-Peaks correspond to the so called A- and B-exciton transition energies. For MoS$_2$ these transition energies at 77~K are 1.84~eV and 2.04~eV, and for MoSe$_2$ at 77~K we find 1.59~eV and 1.80~eV for A- and B-excitons, respectively. As expected, the PL of a MoSSe monolayer lies energetically right in between those of MoS$_2$ and MoSe$_2$, with an A-exciton transition energy of 1.75 eV and a B-exciton of 1.93~eV at 77~K. These values are consistent with prior theoretical predictions \cite{Idrees.2019,RiisJensen.2019}. Therefore, the PL can serve as a further confirmation of the successful transformation of a MoSe$_2$ monolayer into a Janus-type MoSSe monolayer. 
\linebreak

\begin{figure}[htbp]
    \centering
    \includegraphics[width=\textwidth]{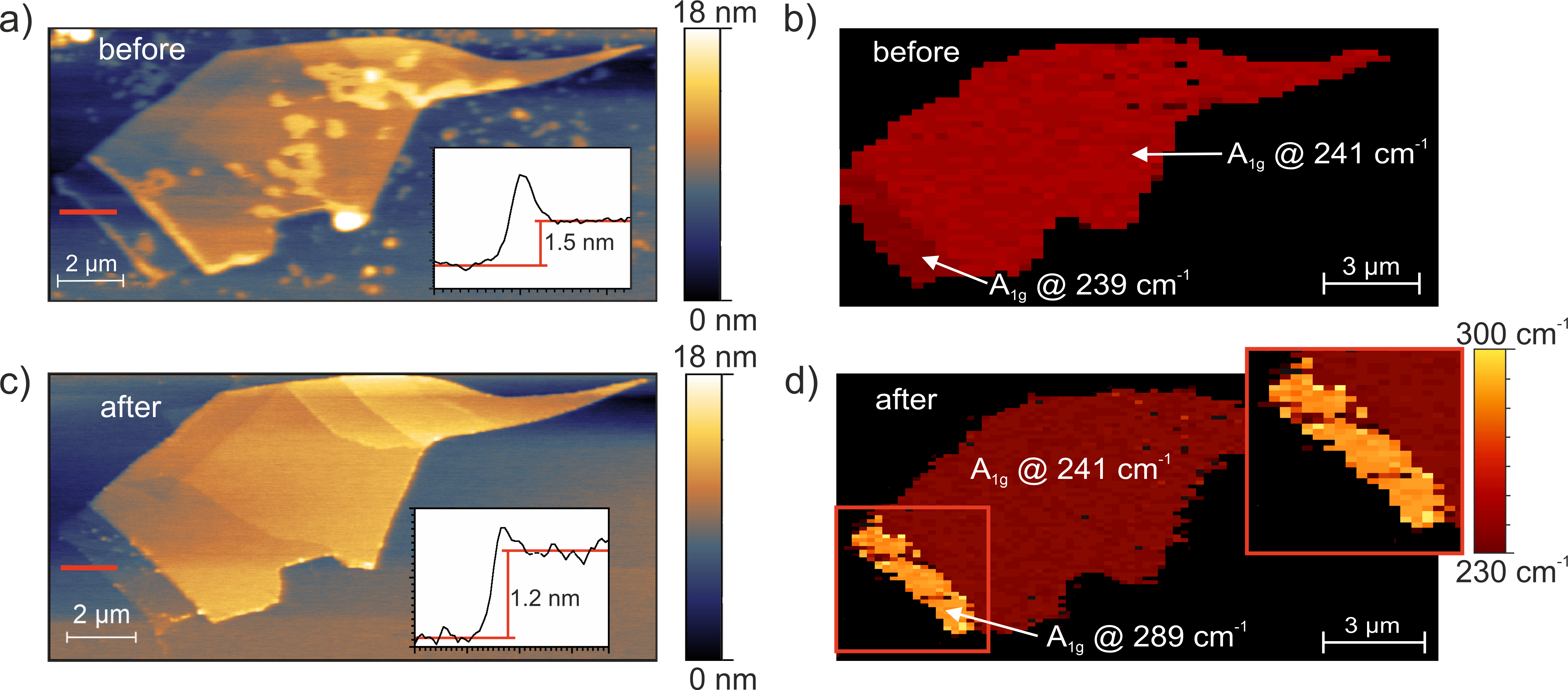}
    \caption{Correlating ambient AFM topography and Raman mappings: (a) and (b) before, and (c) and (d) after the thermal processing. The insets in the AFM images show the height profile of the monolayer at the red lines in the images. The mapping after the process shows inhomogeneities in the monolayer as highlighted in the zoom-in shown in (d). }
    \label{fig:AFM}
\end{figure}

An important quality characteristic of the Janus monolayer after processing is its homogeneity which we will discuss in the following. We observe it via Raman-mappings in direct correlation to atomic force microscopy (AFM) images. The result is shown in Fig. \ref{fig:AFM} before and after a successful process. As seen in the topography, before the thermal sulfurization process, the as-exfoliated flake consists of a thicker part and a monolayer region (see Fig.~\ref{fig:AFM} (a)). This is also reflected in the Raman mapping (b), which shows the typical shift of the A$_{1g}$ out-of-plane vibrational mode to lower wave numbers when thinned down. The images taken before processing show residues from the glue of the tape used for the mechanical exfoliation as well as other adsorbates. The monolayer height as indicated by the inset in the AFM images show a higher-than-nominal step height, which is mainly due to a layer of intercalated water between the substrate and exfoliated flake as well as said adsorbates \cite{Ochedowski.2014b}. These, however, do not appear to have any influence on the Raman signal in the mapping shown in Fig.~\ref{fig:AFM} (b). After the thermal sulfurization process, we observe in the AFM image in Fig.~\ref{fig:AFM} (c) that the flake appears to be significantly cleaner. The adsorbates and glue residues have desorbed after the annealing and therefore the surface of the flake appears much smoother. Also, the step height is reduced by $\sim 300$ pm for the same reason, while still being larger than the nominal monolayer height of 0.7~\AA. This is due to the intercalated water that is  encapsulated by the flake and therefore cannot be fully removed, even when annealed at high temperatures. Previous studies have shown that this can lead to an increased step height on the order of 1~nm \cite{Ochedowski.2014}.

The processing does of course not only affect the morphology but also the materials' composition, which is clearly shown by the Raman mapping presented in Fig. 3(d). Important to note here is, that the mapping highlights only the most dominant mode (excluding the Si-peak), as the scale bar shows. For full spectra taken on multilayer parts, see the supporting information Fig. S1. In Fig. 3(d) we therefore see the very distinctive difference between the monolayer MoSSe and the thicker multilayer part. Due to a stronger signal stemming from the multilayer MoSe$_2$ region, the A$_{1g}$ peak of MoSe$_2$ is also the most dominant one in the thicker parts. However, more interesting is the very homogeneous monolayer region which shows only the Janus-type A$_{1g}$ peak. Correlating this mapping with the AFM image, one can identify the few inhomogeneous spots in the monolayer MoSSe as structural inhomogeneities. For example, the small particles at the lower edge of the monolayer region seen in the topography, correspond roughly to an increased Raman shift in the mapping. These particles could be sulfur residues, which would be plausible due to the fact that gaseous sulfur which is not consumed by the transformation process preferably forms clusters \cite{Raghavachari.1990}.

\begin{figure}[htbp]
    \centering
    \includegraphics[width=\textwidth]{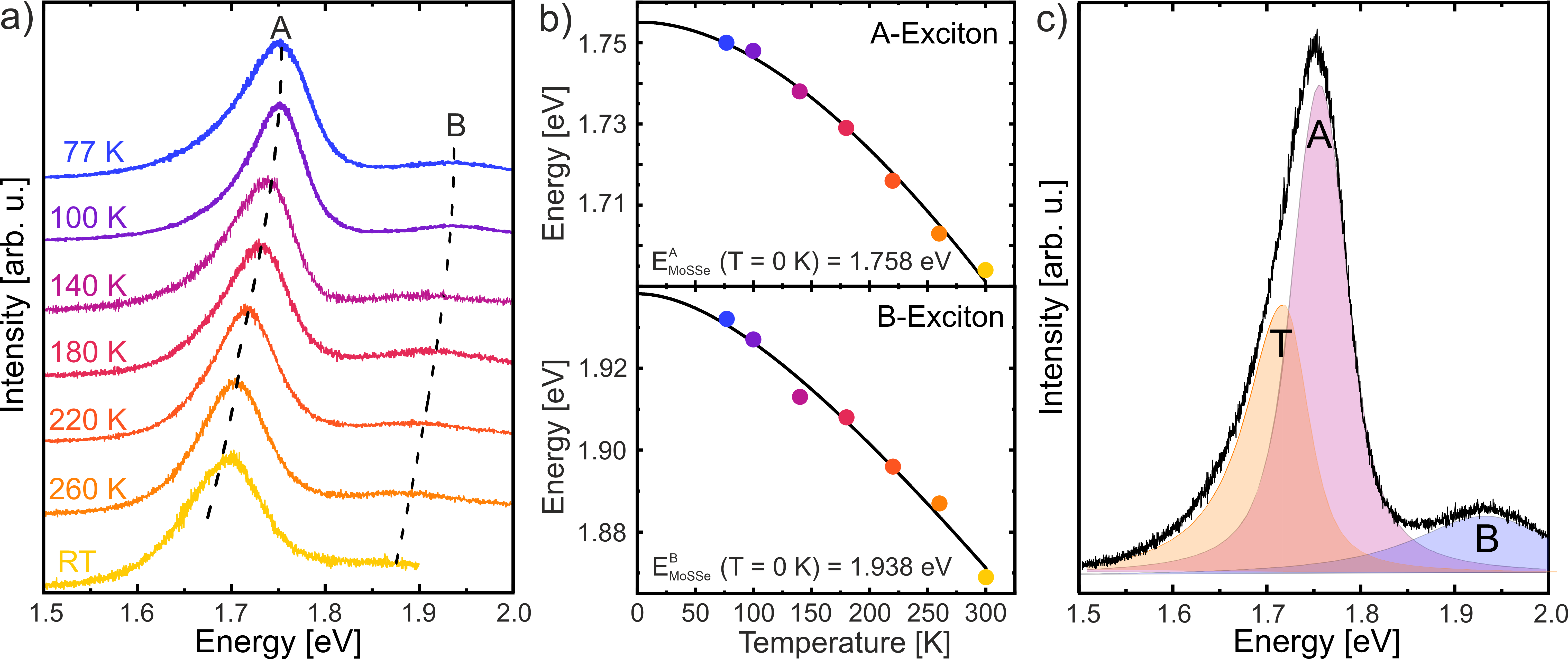}
    \caption{Temperature dependent, normalized photoluminescence spectra of Janus monolayer MoSSe (a) indicating the blue shift of the A and B exciton energies with lower temperatures. Visualized in (b) with Varshni fits for the two excitonic transitions. These indicate \linebreak 0 K transition energies of 1.758 eV and 1.938 eV for A- and B-exciton, respectively.\linebreak (c) Shows the 77 K PL spectra with the A- and B-exciton's, as well as the trion’s (T) fit components.}
    \label{fig:PL-temp}
\end{figure}

Next, we will analyze and discuss the PL spectra of the Janus material in terms of assessing its  quality. Because thermal processes usually obscure the intrinisc mechanisms of the excitonic transitions, low temperature measurements are of key importance for this task. We therefore measured the PL of the Janus-type MoSSe monolayer from room temperature (RT) down to 77 K by cooling with liquid nitrogen. The results are plotted in Fig. \ref{fig:PL-temp} (a). This graph shows for 77 K an A- and a B-exciton at 1.755 eV and 1.921 eV, respectively, as well as a general red shift of these peaks with higher temperatures. The shift of the excitonic peak maxima can be plotted and fitted in excellent agreement with the Varshni relation \cite{Varshni.1967} as seen in Fig. \ref{fig:PL-temp} (b). The Varshni equation is given as
\[
    E(T) = E_0 - \frac{\alpha \cdot T^2}{\beta + T}~,
\]
where E$_0$ [eV] is the excitonic transition energy at 0 K, $\alpha$ [eV/K] is a material specific constant, and $\beta$ [K] corresponds to the Debye temperature $\Theta_{Debye}$.

Therefore, fitting our data with this equation gives us additional, material specific information on our MoSSe monolayer, as well as its excitonic properties. Extrapolating our temperature dependent measurements to 0 K, we find transition energies of 1.758 eV and 1.938 eV for the A- and B-exciton, respectively. This is in good agreement with the theoretical literature \cite{Ji.2018,RiisJensen.2019,Rawat.2020}, however, larger than previously reported experimental results \cite{Trivedi.2020,Petric.2021}. We  believe that this is due to the fact that our base monolayer is exfoliated from a MoSe$_2$ bulk crystal and as such is of better intrinsic quality. The higher transition energies agree with this hypothesis, as a reduced E$_0$ can be correlated to a higher density of defect states in the band gap in general \cite{Hong.2015} and in Janus MoSSe \cite{Long.2021}, specifically. Moreover, most theoretical calculations predict even larger exciton transition energies for perfect Janus MoSSe monolayers under ideal conditions \cite{Idrees.2020,Arra.2019}. We thus conclude, that a trend towards higher energies indicates an improved quality of the material. 

As for the material specific coefficients, we report values of $\alpha$ = 3.78 $\cdot$ 10$^{-4}$~eV/K, and \linebreak
$\beta$ = 214.06~K, or $\alpha$ = 3.94 $\cdot$ 10$^{-4}$~eV/K, and $\beta$ = 230.05~K, as taken from the A- and B-exciton Varshni fits, respectively. The values for $\alpha$ and $\beta$ align well with each other for both excitons. Furthermore, the Debye temperatures of standard TMDCs \cite{Helmrich.2018,Korn.2011}, as well as previous experimental data for MoSSe, \cite{Trivedi.2020} are in good agreement with our data. For a comparison of the temperature dependent behaviour of the base TMDCs MoS$_2$ and MoSe$_2$ with the Janus monolayer, see supporting information Fig. S2.

Lastly, we find a low energy tail of the A-exciton peak in MoSSe, which gets more pronounced with lower temperatures. This is best visualized in Fig. \ref{fig:PL-temp} (c), where the A-exciton peak appears the most asymmetric at 77 K. For monolayer MoS$_2$, this asymmetric tail appears due to trion formation \cite{Christopher.2017}. This seems the most likely case here as well, due to the fact that it appears to have both, the same typical asymmetric distribution, as well as the same correlation with temperature as the trion tail in MoS$_2$, where lower temperatures yield stronger, and more distinct trion peaks \cite{Christopher.2017}. Concluding from the intensive trion formation, the Janus monolayer MoSSe must have excess charge carriers. The intrinsic charge carrier concentration $n_c$ can be approximated by comparing the A-exciton intensity $I_A$ with the charged trion's intensity $I_T$, given according to the mass action model  \cite{Mouri.2013,Ross.2013} as
\[
\frac{I_A}{I_T}\propto \frac{n_c}{k_B T},
\]
with $k_B$ being the Boltzmann constant and $T$ the temperature. For our 77 K PL measurement the intensity ratio is given as $I_A/I_T \simeq 1.61 $, as acquired from the fits' parameters. This is in the same range as the ratio for monolayer MoS$_2$ \cite{Lin.2014}, therefore implying a similar concentration of excess charge carriers. However, many factors such as temperature, pressure, the substrate, the monolayer's strain and intrinsic defects, have an impact on the ratio, thus the comparison to other works must be drawn with reservations. Our own data for MoSe$_2$ and MoS$_2$ monolayers prepared and measured in the exact same manner are shown in the supporting information in Fig. S2. From these we can calculate the $I_A/I_T$ ratios for the 77 K PL measurements in the same manner. For MoSe$_2$ the ratio is the largest with  $I_A/I_T \simeq 5.74$, as the trion's contribution can be barely seen, while the ratio for MoS$_2$ is given as  $I_A/I_T \simeq 2.10$, and thus closer to the value for our MoSSe monolayer. Therefore, we conclude a charge transfer to the Janus monolayer, similar to that of MoS$_2$, which is only really quantifiable at lower temperatures.

\begin{figure}[htbp]
    \centering
    \includegraphics[width=\textwidth]{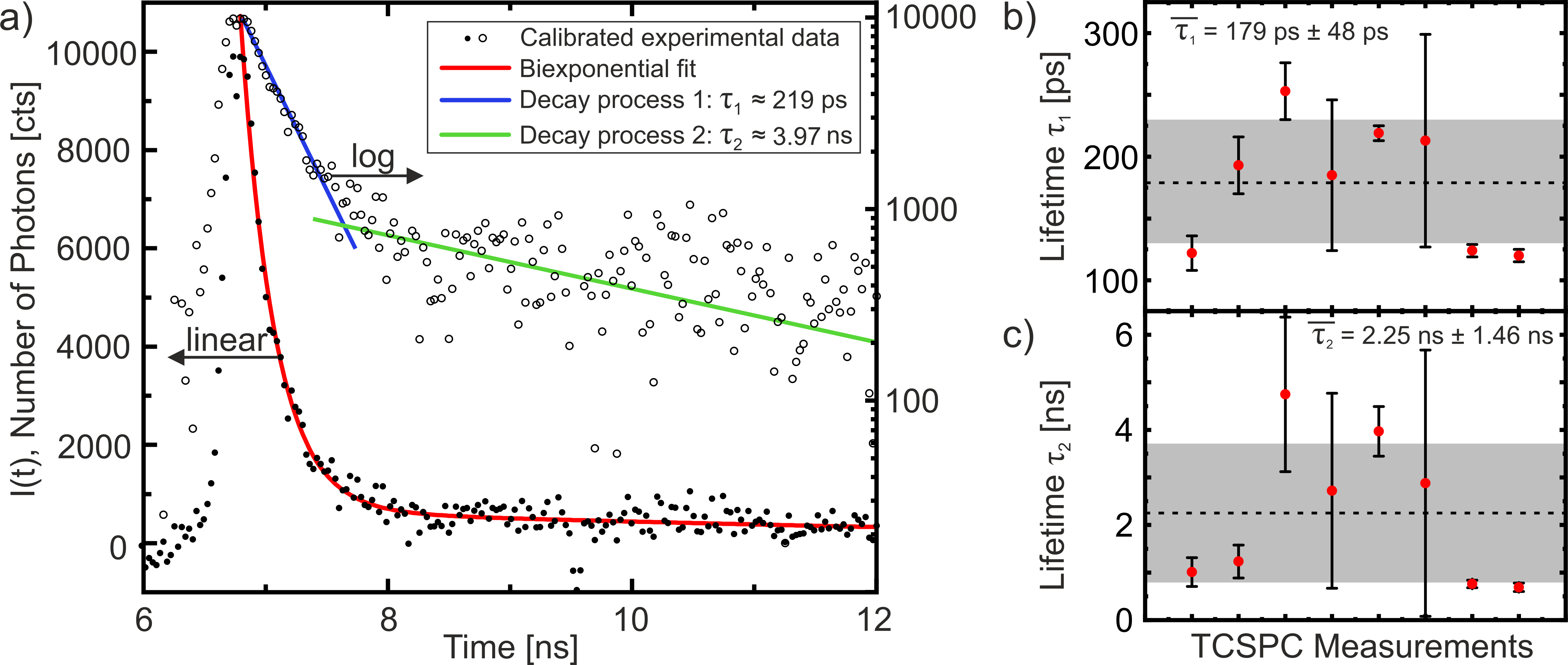}
    \caption{(a) Time-correlated single-photon counting measurement of Janus monolayer MoSSe plotted with a linear (left) and logarithmic (right) scale ordinate. The linear plot shows the bi-exponential fit (red) of the data, while the logarithmic data shows the two separate fit curves. The two different decay processes can be clearly seen, a fast decay (blue) in the range of several hundred picoseconds and a slower decay (green) on the order of a few nanoseconds. In (b) and (c) plots of the statistical distribution of the two lifetime fits stemming from measurements at different positions on monolayer MoSSe are shown.}
    \label{fig:lifetime}
\end{figure}

Finally, we study the exciton dynamics in our Janus material. The recombination processes of excitons and exciton complexes do not happen instantaneously, but on a time scale of a few hundred picoseconds after the excitation with the laser. We measured the lifetime of these processes by time-correlated single-photon counting (TCSPC, for details see method section). A representative calibrated TCSPC measurement at room temperature of monolayer MoSSe is shown in Fig. \ref{fig:lifetime} (a). The calibration spectrum is shown in the supporting information Fig. S3. Here, the data is plotted twice: once with a linear ordinate (left) corresponding to the solid dots, and once on a logarithmic scale (right) corresponding to the open circles. The linearly plotted data shows best that the bi-exponential fits the data well. The general form of the bi-exponential function used to fit the data is given by
\[
    I(t) = I_0 + A_1 \cdot exp\left(-\frac{t-t_0}{\tau_1}\right) + A_2 \cdot exp\left(-\frac{t-t_0}{\tau_2}\right)   ~,
\]
where $I_0$ marks the ordinate offset, $A_i$ are the amplitudes, $t_0$ is the time delay until the exponential decay starts and $\tau_i$ are the cumulative lifetimes of the excited states. The total average lifetime can be calculated from the fit parameters by
\[
\tau_{tot} = \frac{\sum_{i} A_i\cdot (\tau_i)^2}{\sum_{i} A_i\cdot \tau_i}.
\]

In the logarithmic data plot, it becomes more obvious that there are indeed two decay processes at work with different lifetimes. The blue line indicates the short-lived state. Its decay happens on a time scale of $\sim$ 200~ps. While the green line shows a longer lasting decay process on the order of several nanoseconds. This general behaviour of two decay processes with a slower and a faster component is in good agreement with measurements for MoS$_2$ and other TMDCs \cite{Guo.2017b,Wang.2018}. The true nature of the long and short lifetimes are still debated, however it seems clear that the long lifetime is a thermally driven process \cite{Korn.2011,Robert.2016}. At this point it is impossible to separate our data further into A-, B-exciton, and a trion lifetime; the lifetimes we report here are an amalgam of all of those processes including non-radiative recombination and dark excitons.

The average of the two lifetimes measured at different locations on monolayer MoSSe is shown in Fig.~\ref{fig:lifetime} (b) and (c), showing some variation over the flake area. For the corresponding TCSPC measurements see Fig.~S4 and for the fit parameter Tab.~S2 in the supporting information. From this data, we report average lifetimes of $\overline{\tau_1} = $179~ps $\pm$ 48~ps for the fast decay, and $\overline{\tau_2} = $ 2.25~ns $\pm$ 1.46~ns for the slower decay, and an average total lifetime of $\overline{\tau_{tot}}= 1.57$~ns $\pm$ 1.06~ns. These values are in good agreement with those reported for other TMDCs \cite{Guo.2017b,Wang.2018}, however, it appears that they are slightly larger than those reported on average for unmodified MoS$_2$. This could be due to the quantum-confined Stark effect arising from the intrinsic electric field in the Janus monolayer \cite{Dou.2020}, which leads to less overlap of the electron and hole wave-functions, so that recombination occurs on a longer timescale.

\newpage
\section{Conclusion}
In conclusion we have presented a new path to produce Janus-type monolayers and heterostructures (see Supporting Information) by utilizing commercially available, high quality bulk materials for exfoliated samples that were subsequently thermally sulfurized. Our data shows that despite the low intrinsic defect density of the base material the replacement of the upper chalcogen layer is very efficient. We have shown with AFM and correlated Raman mappings, time- and temperature dependent PL spectroscopy, that our Janus MoSSe monolayers are of high quality and we could thus establish corresponding criteria for quality assessment, in particular a high exciton energy. Furthermore, we identified the contribution of a trion to PL spectra recorded at low temperatures, pointing towards a charge transfer to the Janus monolayer. The measured exciton lifetimes are comparable with typical TMDCs, rendering the Janus material equally interesting for applications. We thus see much potential in this type of fabrication method, as new and higher quality base materials have now become accessible, which in turn will lead to better and more variable Janus monolayers and help unlock so far unexplored Janus structures.

\section{Methods}
\subsection{Mechanical Exfoliation}
Our sample fabrication starts with  monolayers mechanically exfoliated from bulk MoSe$_2$. These are prepared onto a silicon substrate with a 385~nm oxide layer after two rounds of cleaning in an ultrasonic bath of laboratory grade acetone and ethanol. We use \textit{Nitto's} \textit{Dicing Tape} to thin down the bulk MoSe$_2$ before pressing it onto the substrate with a heavy steel weight for approximately 30 minutes. 

\subsection{Sulfurization Process}
In each run two exfoliated samples are put inside a multi zone tube furnace approx.~25~cm downstream of a solid sulfur source (85~mg, \textit{Sigma-Aldrich}, 99.98\% sulfur powder). Pure argon gas (99.9\% Ar) is used as an inert carrier gas to evacuate the chamber before (hold 15~min, 200~sccm to clean) and during (constant 75 sccm) the thermal processing. The exfoliated flakes are heated up to 800~$^o$C.  The sulfur is heated with a delay of 9 min up to 150~$^o$C. Both temperatures are held constant for a process time of 20~min. At the end of the process the oven is opened to shock cool the chamber down to room temperature.

\subsection{Raman and Photoluminescence Spectroscopy}
A \textit{WiTec alpha300 RA} confocal Raman spectrometer was used for both Raman and PL measurements. All measurements, including the mappings, were done with a green laser ($\lambda= 532$~nm) with an output power of 0.5~mW. Grid sizes were switched between 600~g/mm for Pl and mappings and the 1800~g/mm grid for single spectra measurements. The low temperature PL measurements were made possible by installing a \textit{Linkam Stage THMS350EV} extension temperature control system to our Raman spectrometer.

\subsection{Atomic Force Microscopy}
The same \textit{WiTec alpha300 RA} confocal Raman spectrometer also includes an AFM setup, which was used for ambient topography measurements directly in correlation to the Raman measurements. The resolution of the images is 512 lines/image $\times$ 512 points/line. The topography was measured with a tip velocity of 1 line/s for forward and backward trace. The cantilevers used for measurement were \textit{NanoSensors' PPP NCHR} extra sharp, reflective coated tips for tapping and non-contact mode.

\subsection{Time-Correlated Single-Photon Counting Measurements}
TCSPC Measurements were done at the same \textit{WiTec alpha300 RA} confocal Raman spectrometer, by using a \textit{PicoQuant LDH-D-C-405} pulsed laser with a wavelength of 405~nm and a pulse width of $< 50$~ps. For the single-photon detection an avalanche photodiode (resolution: $< 50$~ps, range: 400--1050~nm, dark count rate: $< 50$~cts/s) and a time to digital module (readout speed: $< 80$~Hz) were used. Calibration of the measurement was done in the dark, beforehand. For a representative calibration spectrum see the supporting information. Measurements were done at RT with a repetition frequency of 40 MHz, over a 25 s integration time on a time scale from 0 to 25~ns (1024 time bins), yielding a resolution of 30~ps per time bin. Each final spectrum (as the one shown) was then averaged over 30 single spectra.

%%%%%%%%%%%%%%%%%%%%%%%%Acknowledgement%%%%%%%%%%%%%%%%%%%%%%%%%%%%%%%

\begin{acknowledgement}
We thank P. Kratzer for fruitful discussions. We thank the faculty of physics for a seeding grant and acknowledge financial support from the German Research Foundation (DFG) by funding SCHL 384/20-1
(project number 406129719) and project C5 within the SFB1242 “Non-Equilibrium Dynamics of Condensed Matter in the Time Domain” (project number 278162697).

\end{acknowledgement}

%%%%%%%%%%%%%%%%%%%%%%%%SupportingInformation%%%%%%%%%%%%%%%%%%%%%%%%%
\begin{suppinfo}
Supporting Information is available.\\
Supplementary to the data shown here, we present Raman spectra of a multilayer MoSe$_2$ part of a sample after the processing. This gives additional information on how the process cuts and replaces the Se-bonds and shows an as-processed heterostructure of a monolayer MoSSe on top of few-layer MoSe$_2$. Moreover, we present a more detailed comparison of the temperature-dependent PL of MoSSe in comparison with MoSe$_2$ and MoS$_2$, both of which show realistic values for the parameters of the respective Varshni-fits, corroborating our data for the Janus MoSSe monolayer further. Lastly, we present a calibration spectrum for the TCSPC measurement and explain some of its follies, as well as spectra of the other TCSPC measurements from different points on monolayer MoSSe.    
\end{suppinfo}

%%%%%%%%%%%%%%%%%%%%%%%%%%Bibliography%%%%%%%%%%%%%%%%%%%%%%%%%%%%%%%%%
\bibliography{Lifetime_MoSSe_paper.bib}

\providecommand{\latin}[1]{#1}
\makeatletter
\providecommand{\doi}
  {\begingroup\let\do\@makeother\dospecials
  \catcode`\{=1 \catcode`\}=2 \doi@aux}
\providecommand{\doi@aux}[1]{\endgroup\texttt{#1}}
\makeatother
\providecommand*\mcitethebibliography{\thebibliography}
\csname @ifundefined\endcsname{endmcitethebibliography}
  {\let\endmcitethebibliography\endthebibliography}{}
\begin{mcitethebibliography}{44}
\providecommand*\natexlab[1]{#1}
\providecommand*\mciteSetBstSublistMode[1]{}
\providecommand*\mciteSetBstMaxWidthForm[2]{}
\providecommand*\mciteBstWouldAddEndPuncttrue
  {\def\EndOfBibitem{\unskip.}}
\providecommand*\mciteBstWouldAddEndPunctfalse
  {\let\EndOfBibitem\relax}
\providecommand*\mciteSetBstMidEndSepPunct[3]{}
\providecommand*\mciteSetBstSublistLabelBeginEnd[3]{}
\providecommand*\EndOfBibitem{}
\mciteSetBstSublistMode{f}
\mciteSetBstMaxWidthForm{subitem}{(\alph{mcitesubitemcount})}
\mciteSetBstSublistLabelBeginEnd
  {\mcitemaxwidthsubitemform\space}
  {\relax}
  {\relax}

\bibitem[Lu \latin{et~al.}(2017)Lu, Zhu, Xiao, Chuu, Han, Chiu, Cheng, Yang,
  Wei, Yang, Wang, Sokaras, Nordlund, Yang, Muller, Chou, Zhang, and
  Li]{Lu.2017}
Lu,~A.-Y. \latin{et~al.}  {Janus monolayers of transition metal
  dichalcogenides}. \emph{{Nature Nanotechnology}} \textbf{2017}, \emph{12},
  744--749\relax
\mciteBstWouldAddEndPuncttrue
\mciteSetBstMidEndSepPunct{\mcitedefaultmidpunct}
{\mcitedefaultendpunct}{\mcitedefaultseppunct}\relax
\EndOfBibitem
\bibitem[Zhang \latin{et~al.}(2017)Zhang, Jia, Kholmanov, Dong, Er, Chen, Guo,
  Jin, Shenoy, Shi, and Lou]{Zhang.2017}
Zhang,~J.; Jia,~S.; Kholmanov,~I.; Dong,~L.; Er,~D.; Chen,~W.; Guo,~H.;
  Jin,~Z.; Shenoy,~V.~B.; Shi,~L.; Lou,~J. {Janus Monolayer Transition-Metal
  Dichalcogenides}. \emph{{ACS Nano}} \textbf{2017}, \emph{11},
  8192--8198\relax
\mciteBstWouldAddEndPuncttrue
\mciteSetBstMidEndSepPunct{\mcitedefaultmidpunct}
{\mcitedefaultendpunct}{\mcitedefaultseppunct}\relax
\EndOfBibitem
\bibitem[Sant \latin{et~al.}(2020)Sant, Gay, Marty, Lisi, Harrabi, Vergnaud,
  Dau, Weng, Coraux, Gauthier, Renault, Renaud, and Jamet]{Sant.2020}
Sant,~R.; Gay,~M.; Marty,~A.; Lisi,~S.; Harrabi,~R.; Vergnaud,~C.; Dau,~M.~T.;
  Weng,~X.; Coraux,~J.; Gauthier,~N.; Renault,~O.; Renaud,~G.; Jamet,~M.
  {Synthesis of epitaxial monolayer Janus S{P}t{S}e}. \emph{{npj 2D Materials
  and Applications}} \textbf{2020}, \emph{4}\relax
\mciteBstWouldAddEndPuncttrue
\mciteSetBstMidEndSepPunct{\mcitedefaultmidpunct}
{\mcitedefaultendpunct}{\mcitedefaultseppunct}\relax
\EndOfBibitem
\bibitem[Li \latin{et~al.}(2018)Li, Cheng, and Huang]{Li.2018}
Li,~R.; Cheng,~Y.; Huang,~W. {Recent Progress of Janus 2D Transition Metal
  Chalcogenides: From Theory to Experiments}. \emph{{Small}} \textbf{2018},
  \emph{14}, e1802091\relax
\mciteBstWouldAddEndPuncttrue
\mciteSetBstMidEndSepPunct{\mcitedefaultmidpunct}
{\mcitedefaultendpunct}{\mcitedefaultseppunct}\relax
\EndOfBibitem
\bibitem[Zhang \latin{et~al.}(2020)Zhang, Yang, Gong, Pan, Wang, Guo, Zhang,
  and Fu]{Zhang.2020}
Zhang,~L.; Yang,~Z.; Gong,~T.; Pan,~R.; Wang,~H.; Guo,~Z.; Zhang,~H.; Fu,~X.
  {Recent advances in emerging Janus two-dimensional materials: from
  fundamental physics to device applications}. \emph{{J. Mat. Chem. A}}
  \textbf{2020}, \emph{8}, 8813--8830\relax
\mciteBstWouldAddEndPuncttrue
\mciteSetBstMidEndSepPunct{\mcitedefaultmidpunct}
{\mcitedefaultendpunct}{\mcitedefaultseppunct}\relax
\EndOfBibitem
\bibitem[Wei \latin{et~al.}(2020)Wei, Tang, Shang, Ju, and Kou]{Wei.2020}
Wei,~Y.; Tang,~X.; Shang,~J.; Ju,~L.; Kou,~L. {Two-dimensional functional
  materials: from properties to potential applications}. \emph{{International
  Journal of Smart and Nano Materials}} \textbf{2020}, \emph{11},
  247--264\relax
\mciteBstWouldAddEndPuncttrue
\mciteSetBstMidEndSepPunct{\mcitedefaultmidpunct}
{\mcitedefaultendpunct}{\mcitedefaultseppunct}\relax
\EndOfBibitem
\bibitem[Riis-Jensen \latin{et~al.}(2019)Riis-Jensen, Deilmann, Olsen, and
  Thygesen]{RiisJensen.2019}
Riis-Jensen,~A.~C.; Deilmann,~T.; Olsen,~T.; Thygesen,~K.~S. {Classifying the
  Electronic and Optical Properties of Janus Monolayers}. \emph{{ACS Nano}}
  \textbf{2019}, \emph{13}, 13354--13364\relax
\mciteBstWouldAddEndPuncttrue
\mciteSetBstMidEndSepPunct{\mcitedefaultmidpunct}
{\mcitedefaultendpunct}{\mcitedefaultseppunct}\relax
\EndOfBibitem
\bibitem[Dong \latin{et~al.}(2017)Dong, Lou, and Shenoy]{Dong.2017b}
Dong,~L.; Lou,~J.; Shenoy,~V.~B. {Large In-Plane and Vertical Piezoelectricity
  in Janus Transition Metal Dichalchogenides}. \emph{{ACS Nano}} \textbf{2017},
  \emph{11}, 8242--8248\relax
\mciteBstWouldAddEndPuncttrue
\mciteSetBstMidEndSepPunct{\mcitedefaultmidpunct}
{\mcitedefaultendpunct}{\mcitedefaultseppunct}\relax
\EndOfBibitem
\bibitem[Ahammed \latin{et~al.}(2020)Ahammed, Jena, Rawat, Mohanta, Dimple, and
  de~Sarkar]{Ahammed.2020}
Ahammed,~R.; Jena,~N.; Rawat,~A.; Mohanta,~M.~K.; Dimple,; de~Sarkar,~A.
  {Ultrahigh Out-of-Plane Piezoelectricity Meets Giant Rashba Effect in 2D
  Janus Monolayers and Bilayers of Group IV Transition-Metal Trichalcogenides}.
  \emph{{J. Phys. Chem. C}} \textbf{2020}, \emph{124}, 21250--21260\relax
\mciteBstWouldAddEndPuncttrue
\mciteSetBstMidEndSepPunct{\mcitedefaultmidpunct}
{\mcitedefaultendpunct}{\mcitedefaultseppunct}\relax
\EndOfBibitem
\bibitem[Rawat \latin{et~al.}(2020)Rawat, Mohanta, Jena, Dimple, Ahammed, and
  de~Sarkar]{Rawat.2020}
Rawat,~A.; Mohanta,~M.~K.; Jena,~N.; Dimple,; Ahammed,~R.; de~Sarkar,~A.
  {Nanoscale Interfaces of Janus Monolayers of Transition Metal Dichalcogenides
  for 2D Photovoltaic and Piezoelectric Applications}. \emph{{J. Phys. Chem.
  C}} \textbf{2020}, \emph{124}, 10385--10397\relax
\mciteBstWouldAddEndPuncttrue
\mciteSetBstMidEndSepPunct{\mcitedefaultmidpunct}
{\mcitedefaultendpunct}{\mcitedefaultseppunct}\relax
\EndOfBibitem
\bibitem[Ji \latin{et~al.}(2018)Ji, Yang, Lin, Hou, Wang, Li, and Lee]{Ji.2018}
Ji,~Y.; Yang,~M.; Lin,~H.; Hou,~T.; Wang,~L.; Li,~Y.; Lee,~S.-T. {Janus
  Structures of Transition Metal Dichalcogenides as the Heterojunction
  Photocatalysts for Water Splitting}. \emph{{J. Phys. Chem. C}} \textbf{2018},
  \emph{122}, 3123--3129\relax
\mciteBstWouldAddEndPuncttrue
\mciteSetBstMidEndSepPunct{\mcitedefaultmidpunct}
{\mcitedefaultendpunct}{\mcitedefaultseppunct}\relax
\EndOfBibitem
\bibitem[Din \latin{et~al.}(2019)Din, Idrees, Albar, Shafiq, Ahmad, Nguyen, and
  Amin]{Din.2019}
Din,~H.~U.; Idrees,~M.; Albar,~A.; Shafiq,~M.; Ahmad,~I.; Nguyen,~C.~V.;
  Amin,~B. {Rashba spin splitting and photocatalytic properties of
  GeC$-$M{S}{S}e ( M=Mo , W) van der Waals heterostructures}. \emph{{Phys. Rev.
  B}} \textbf{2019}, \emph{100}\relax
\mciteBstWouldAddEndPuncttrue
\mciteSetBstMidEndSepPunct{\mcitedefaultmidpunct}
{\mcitedefaultendpunct}{\mcitedefaultseppunct}\relax
\EndOfBibitem
\bibitem[Idrees \latin{et~al.}(2019)Idrees, Din, Ali, Rehman, Hussain, Nguyen,
  Ahmad, and Amin]{Idrees.2019}
Idrees,~M.; Din,~H.~U.; Ali,~R.; Rehman,~G.; Hussain,~T.; Nguyen,~C.~V.;
  Ahmad,~I.; Amin,~B. {Optoelectronic and solar cell applications of Janus
  monolayers and their van der Waals heterostructures}. \emph{{Phys. Chem.
  Chem. Phys.}} \textbf{2019}, \emph{21}, 18612--18621\relax
\mciteBstWouldAddEndPuncttrue
\mciteSetBstMidEndSepPunct{\mcitedefaultmidpunct}
{\mcitedefaultendpunct}{\mcitedefaultseppunct}\relax
\EndOfBibitem
\bibitem[Cheng \latin{et~al.}(2013)Cheng, Zhu, Tahir, and
  Schwingenschl{\"o}gl]{Cheng.2013}
Cheng,~Y.~C.; Zhu,~Z.~Y.; Tahir,~M.; Schwingenschl{\"o}gl,~U.
  {Spin-orbit--induced spin splittings in polar transition metal dichalcogenide
  monolayers}. \emph{{EPL}} \textbf{2013}, \emph{102}, 57001\relax
\mciteBstWouldAddEndPuncttrue
\mciteSetBstMidEndSepPunct{\mcitedefaultmidpunct}
{\mcitedefaultendpunct}{\mcitedefaultseppunct}\relax
\EndOfBibitem
\bibitem[Li \latin{et~al.}(2020)Li, Qin, Ko, Trivedi, Hajra, Sayyad, Liu, Shim,
  Zhuang, and Tongay]{Li.2020}
Li,~H.; Qin,~Y.; Ko,~B.; Trivedi,~D.~B.; Hajra,~D.; Sayyad,~M.~Y.; Liu,~L.;
  Shim,~S.-H.; Zhuang,~H.; Tongay,~S. {Anomalous Behavior of 2D Janus Excitonic
  Layers under Extreme Pressures}. \emph{{Adv. Mat.}} \textbf{2020}, \emph{32},
  e2002401\relax
\mciteBstWouldAddEndPuncttrue
\mciteSetBstMidEndSepPunct{\mcitedefaultmidpunct}
{\mcitedefaultendpunct}{\mcitedefaultseppunct}\relax
\EndOfBibitem
\bibitem[Trivedi \latin{et~al.}(2020)Trivedi, Turgut, Qin, Sayyad, Hajra,
  Howell, Liu, Yang, Patoary, Li, Petri{\'c}, Meyer, Kremser, Barbone, Soavi,
  Stier, M{\"u}ller, Yang, Esqueda, Zhuang, Finley, and Tongay]{Trivedi.2020}
Trivedi,~D.~B. \latin{et~al.}  {Room-Temperature Synthesis of 2D Janus Crystals
  and their Heterostructures}. \emph{{Adv. Mat.}} \textbf{2020}, \emph{32},
  e2006320\relax
\mciteBstWouldAddEndPuncttrue
\mciteSetBstMidEndSepPunct{\mcitedefaultmidpunct}
{\mcitedefaultendpunct}{\mcitedefaultseppunct}\relax
\EndOfBibitem
\bibitem[Zhang \latin{et~al.}(2020)Zhang, Guo, Ji, Lu, Su, Wang, Puretzky,
  Geohegan, Qian, Fang, Kaxiras, Kong, and Huang]{Zhang.2020b}
Zhang,~K.; Guo,~Y.; Ji,~Q.; Lu,~A.-Y.; Su,~C.; Wang,~H.; Puretzky,~A.~A.;
  Geohegan,~D.~B.; Qian,~X.; Fang,~S.; Kaxiras,~E.; Kong,~J.; Huang,~S.
  {Enhancement of van der Waals Interlayer Coupling through Polar Janus
  Mo{S}{S}e}. \emph{{J. Am. Chem. Soc.}} \textbf{2020}, \emph{142},
  17499--17507\relax
\mciteBstWouldAddEndPuncttrue
\mciteSetBstMidEndSepPunct{\mcitedefaultmidpunct}
{\mcitedefaultendpunct}{\mcitedefaultseppunct}\relax
\EndOfBibitem
\bibitem[Petri{\'c} \latin{et~al.}(2021)Petri{\'c}, Kremser, Barbone, Qin,
  Sayyad, Shen, Tongay, Finley, Botello-M{\'e}ndez, and
  M{\"u}ller]{Petric.2021}
Petri{\'c},~M.~M.; Kremser,~M.; Barbone,~M.; Qin,~Y.; Sayyad,~Y.; Shen,~Y.;
  Tongay,~S.; Finley,~J.~J.; Botello-M{\'e}ndez,~A.~R.; M{\"u}ller,~K. {Raman
  spectrum of Janus transition metal dichalcogenide monolayers W{S}{S}e and
  Mo{S}{S}e}. \emph{{Phys. Rev. B}} \textbf{2021}, \emph{103}\relax
\mciteBstWouldAddEndPuncttrue
\mciteSetBstMidEndSepPunct{\mcitedefaultmidpunct}
{\mcitedefaultendpunct}{\mcitedefaultseppunct}\relax
\EndOfBibitem
\bibitem[Duong \latin{et~al.}(2017)Duong, Yun, and Lee]{Duong.2017}
Duong,~D.~L.; Yun,~S.~J.; Lee,~Y.~H. {van der Waals Layered Materials:
  Opportunities and Challenges}. \emph{{ACS Nano}} \textbf{2017}, \emph{11},
  11803--11830\relax
\mciteBstWouldAddEndPuncttrue
\mciteSetBstMidEndSepPunct{\mcitedefaultmidpunct}
{\mcitedefaultendpunct}{\mcitedefaultseppunct}\relax
\EndOfBibitem
\bibitem[Mounet \latin{et~al.}(2018)Mounet, Schwaller, and
  et~al.]{Marzari.2018}
Mounet,~M.,~N.and~Gibertini; Schwaller,~P.; et~al., {Two-dimensional materials
  from high-throughput computational exfoliation of experimentally known
  compounds.} \emph{{Nature Nanotech.}} \textbf{2018}, \emph{13},
  246--252\relax
\mciteBstWouldAddEndPuncttrue
\mciteSetBstMidEndSepPunct{\mcitedefaultmidpunct}
{\mcitedefaultendpunct}{\mcitedefaultseppunct}\relax
\EndOfBibitem
\bibitem[Frisenda \latin{et~al.}(2020)Frisenda, Niu, Gant, Mu{\~n}oz, and
  Castellanos-Gomez]{Frisenda.2020}
Frisenda,~R.; Niu,~Y.; Gant,~P.; Mu{\~n}oz,~M.; Castellanos-Gomez,~A.
  {Naturally occurring van der Waals materials}. \emph{{npj 2D Materials and
  Applications}} \textbf{2020}, \emph{4}, 1--13\relax
\mciteBstWouldAddEndPuncttrue
\mciteSetBstMidEndSepPunct{\mcitedefaultmidpunct}
{\mcitedefaultendpunct}{\mcitedefaultseppunct}\relax
\EndOfBibitem
\bibitem[Hong \latin{et~al.}(2015)Hong, Hu, Probert, Li, Lv, Yang, Gu, Mao,
  Feng, Xie, Zhang, Wu, Zhang, Jin, Ji, Zhang, Yuan, and Zhang]{Hong.2015}
Hong,~J. \latin{et~al.}  {Exploring atomic defects in molybdenum disulphide
  monolayers}. \emph{{Nature Communications}} \textbf{2015}, \emph{6},
  6293\relax
\mciteBstWouldAddEndPuncttrue
\mciteSetBstMidEndSepPunct{\mcitedefaultmidpunct}
{\mcitedefaultendpunct}{\mcitedefaultseppunct}\relax
\EndOfBibitem
\bibitem[Pollmann \latin{et~al.}(2018)Pollmann, Madau{\ss}, Zeuner, and
  Schleberger]{Pollmann.2018}
Pollmann,~E.; Madau{\ss},~L.; Zeuner,~V.; Schleberger,~M. In \emph{{Enc.
  Interf. Chem.}}; Wandelt,~K., Ed.; Elsevier: Amsterdam, Netherlands and
  Oxford, UK and Cambidge, USA, 2018; pp 338--343\relax
\mciteBstWouldAddEndPuncttrue
\mciteSetBstMidEndSepPunct{\mcitedefaultmidpunct}
{\mcitedefaultendpunct}{\mcitedefaultseppunct}\relax
\EndOfBibitem
\bibitem[Pollmann \latin{et~al.}(2020)Pollmann, Madau{\ss}, Schumacher, Kumar,
  Heuvel, {vom Ende}, Yilmaz, G{\"u}ng{\"o}rm{\"u}s, and
  Schleberger]{Pollmann.2020}
Pollmann,~E.; Madau{\ss},~L.; Schumacher,~S.; Kumar,~U.; Heuvel,~F.; {vom
  Ende},~C.; Yilmaz,~S.; G{\"u}ng{\"o}rm{\"u}s,~S.; Schleberger,~M. {Apparent
  differences between single layer molybdenum disulphide fabricated via
  chemical vapour deposition and exfoliation}. \emph{{Nanotechnology}}
  \textbf{2020}, \emph{31}, 505604\relax
\mciteBstWouldAddEndPuncttrue
\mciteSetBstMidEndSepPunct{\mcitedefaultmidpunct}
{\mcitedefaultendpunct}{\mcitedefaultseppunct}\relax
\EndOfBibitem
\bibitem[Chu \latin{et~al.}(2016)Chu, Park, and Shen]{Chu.2016}
Chu,~S.; Park,~C.; Shen,~G. {Structural characteristic correlated to the
  electronic band gap in Mo{S}$_2$}. \emph{{Phys. Rev. B}} \textbf{2016},
  \emph{94}\relax
\mciteBstWouldAddEndPuncttrue
\mciteSetBstMidEndSepPunct{\mcitedefaultmidpunct}
{\mcitedefaultendpunct}{\mcitedefaultseppunct}\relax
\EndOfBibitem
\bibitem[Xia \latin{et~al.}(2018)Xia, Xiong, Du, Wang, Peng, and Li]{Xia.2018}
Xia,~C.; Xiong,~W.; Du,~J.; Wang,~T.; Peng,~Y.; Li,~J. {Universality of
  electronic characteristics and photocatalyst applications in the
  two-dimensional Janus transition metal dichalcogenides}. \emph{{Phys. Rev.
  B}} \textbf{2018}, \emph{98}\relax
\mciteBstWouldAddEndPuncttrue
\mciteSetBstMidEndSepPunct{\mcitedefaultmidpunct}
{\mcitedefaultendpunct}{\mcitedefaultseppunct}\relax
\EndOfBibitem
\bibitem[Ochedowski \latin{et~al.}(2014)Ochedowski, Marinov, Scheuschner,
  Poloczek, Bussmann, Maultzsch, and Schleberger]{Ochedowski.2014b}
Ochedowski,~O.; Marinov,~K.; Scheuschner,~N.; Poloczek,~A.; Bussmann,~B.~K.;
  Maultzsch,~J.; Schleberger,~M. {Effect of contaminations and surface
  preparation on the work function of single layer Mo{S}$_2$}. \emph{{Beilst.
  J. Nanotech.}} \textbf{2014}, \emph{5}, 291--297\relax
\mciteBstWouldAddEndPuncttrue
\mciteSetBstMidEndSepPunct{\mcitedefaultmidpunct}
{\mcitedefaultendpunct}{\mcitedefaultseppunct}\relax
\EndOfBibitem
\bibitem[Ochedowski \latin{et~al.}(2014)Ochedowski, Bussmann, and
  Schleberger]{Ochedowski.2014}
Ochedowski,~O.; Bussmann,~B.~K.; Schleberger,~M. {Graphene on mica -
  intercalated water trapped for life}. \emph{{Sci. Rep.}} \textbf{2014},
  \emph{4}, 6003\relax
\mciteBstWouldAddEndPuncttrue
\mciteSetBstMidEndSepPunct{\mcitedefaultmidpunct}
{\mcitedefaultendpunct}{\mcitedefaultseppunct}\relax
\EndOfBibitem
\bibitem[Raghavachari \latin{et~al.}(1990)Raghavachari, Rohlfing, and
  Binkley]{Raghavachari.1990}
Raghavachari,~K.; Rohlfing,~C.~M.; Binkley,~J.~S. {Structures and stabilities
  of sulfur clusters}. \emph{{J. Chem. Phys.}} \textbf{1990}, \emph{93},
  5862--5874\relax
\mciteBstWouldAddEndPuncttrue
\mciteSetBstMidEndSepPunct{\mcitedefaultmidpunct}
{\mcitedefaultendpunct}{\mcitedefaultseppunct}\relax
\EndOfBibitem
\bibitem[Varshni(1967)]{Varshni.1967}
Varshni,~Y.~P. {Temperature dependence of the energy gap in semiconductors}.
  \emph{{Physica}} \textbf{1967}, \emph{34}, 149--154\relax
\mciteBstWouldAddEndPuncttrue
\mciteSetBstMidEndSepPunct{\mcitedefaultmidpunct}
{\mcitedefaultendpunct}{\mcitedefaultseppunct}\relax
\EndOfBibitem
\bibitem[Long \latin{et~al.}(2021)Long, Dai, and Jin]{Long.2021}
Long,~C.; Dai,~Y.; Jin,~H. {Effect of point defects on electronic and excitonic
  properties in Janus-Mo{S}{S}e monolayer}. \emph{{Phys. Rev. B}}
  \textbf{2021}, \emph{104}\relax
\mciteBstWouldAddEndPuncttrue
\mciteSetBstMidEndSepPunct{\mcitedefaultmidpunct}
{\mcitedefaultendpunct}{\mcitedefaultseppunct}\relax
\EndOfBibitem
\bibitem[Idrees \latin{et~al.}(2020)Idrees, Din, Rehman, Shafiq, Saeed, Bui,
  Nguyen, and Amin]{Idrees.2020}
Idrees,~M.; Din,~H.~U.; Rehman,~S.~U.; Shafiq,~M.; Saeed,~Y.; Bui,~H.~D.;
  Nguyen,~C.~V.; Amin,~B. {Electronic properties and enhanced photocatalytic
  performance of van der Waals heterostructures of ZnO and Janus transition
  metal dichalcogenides}. \emph{{Phys. Chem. Chem. Phys.}} \textbf{2020},
  \emph{22}, 10351--10359\relax
\mciteBstWouldAddEndPuncttrue
\mciteSetBstMidEndSepPunct{\mcitedefaultmidpunct}
{\mcitedefaultendpunct}{\mcitedefaultseppunct}\relax
\EndOfBibitem
\bibitem[Arra \latin{et~al.}(2019)Arra, Babar, and Kabir]{Arra.2019}
Arra,~S.; Babar,~R.; Kabir,~M. {van der Waals heterostructure for
  photocatalysis: Graphitic carbon nitride and Janus transition-metal
  dichalcogenides}. \emph{{Phys. Rev. Mat.}} \textbf{2019}, \emph{3}\relax
\mciteBstWouldAddEndPuncttrue
\mciteSetBstMidEndSepPunct{\mcitedefaultmidpunct}
{\mcitedefaultendpunct}{\mcitedefaultseppunct}\relax
\EndOfBibitem
\bibitem[Helmrich \latin{et~al.}(2018)Helmrich, Schneider, Achtstein, Arora,
  Herzog, de~Vasconcellos, Kolarczik, Sch{\"o}ps, Bratschitsch, Woggon, and
  Owschimikow]{Helmrich.2018}
Helmrich,~S.; Schneider,~R.; Achtstein,~A.~W.; Arora,~A.; Herzog,~B.;
  de~Vasconcellos,~S.~M.; Kolarczik,~M.; Sch{\"o}ps,~O.; Bratschitsch,~R.;
  Woggon,~U.; Owschimikow,~N. {Exciton--phonon coupling in mono- and bilayer
  MoTe 2}. \emph{{2D Materials}} \textbf{2018}, \emph{5}, 045007\relax
\mciteBstWouldAddEndPuncttrue
\mciteSetBstMidEndSepPunct{\mcitedefaultmidpunct}
{\mcitedefaultendpunct}{\mcitedefaultseppunct}\relax
\EndOfBibitem
\bibitem[Korn \latin{et~al.}(2011)Korn, Heydrich, Hirmer, Schmutzler, and
  Sch{\"u}ller]{Korn.2011}
Korn,~T.; Heydrich,~S.; Hirmer,~M.; Schmutzler,~J.; Sch{\"u}ller,~C.
  {Low-temperature photocarrier dynamics in monolayer Mo{S}$_2$}. \emph{{Appl.
  Phys. Lett.}} \textbf{2011}, \emph{99}, 102109\relax
\mciteBstWouldAddEndPuncttrue
\mciteSetBstMidEndSepPunct{\mcitedefaultmidpunct}
{\mcitedefaultendpunct}{\mcitedefaultseppunct}\relax
\EndOfBibitem
\bibitem[Christopher \latin{et~al.}(2017)Christopher, Goldberg, and
  Swan]{Christopher.2017}
Christopher,~J.~W.; Goldberg,~B.~B.; Swan,~A.~K. {Long tailed trions in
  monolayer Mo{S}$_2$: Temperature dependent asymmetry and resulting red-shift
  of trion photoluminescence spectra}. \emph{{Sci. Rep.}} \textbf{2017},
  \emph{7}, 14062\relax
\mciteBstWouldAddEndPuncttrue
\mciteSetBstMidEndSepPunct{\mcitedefaultmidpunct}
{\mcitedefaultendpunct}{\mcitedefaultseppunct}\relax
\EndOfBibitem
\bibitem[Mouri \latin{et~al.}(2013)Mouri, Miyauchi, and Matsuda]{Mouri.2013}
Mouri,~S.; Miyauchi,~Y.; Matsuda,~K. {Tunable photoluminescence of monolayer
  Mo{S}$_2$ via chemical doping}. \emph{{Nano Letters}} \textbf{2013},
  \emph{13}, 5944--5948\relax
\mciteBstWouldAddEndPuncttrue
\mciteSetBstMidEndSepPunct{\mcitedefaultmidpunct}
{\mcitedefaultendpunct}{\mcitedefaultseppunct}\relax
\EndOfBibitem
\bibitem[Ross \latin{et~al.}(2013)Ross, Wu, Yu, Ghimire, Jones, Aivazian, Yan,
  Mandrus, {Di Xiao}, Yao, and Xu]{Ross.2013}
Ross,~J.~S.; Wu,~S.; Yu,~H.; Ghimire,~N.~J.; Jones,~A.~M.; Aivazian,~G.;
  Yan,~J.; Mandrus,~D.~G.; {Di Xiao},; Yao,~W.; Xu,~X. {Electrical control of
  neutral and charged excitons in a monolayer semiconductor}. \emph{{Nature
  Comm.}} \textbf{2013}, \emph{4}, 1474\relax
\mciteBstWouldAddEndPuncttrue
\mciteSetBstMidEndSepPunct{\mcitedefaultmidpunct}
{\mcitedefaultendpunct}{\mcitedefaultseppunct}\relax
\EndOfBibitem
\bibitem[Lin \latin{et~al.}(2014)Lin, Ling, Yu, Huang, Hsu, Lee, Kong,
  Dresselhaus, and Palacios]{Lin.2014}
Lin,~Y.; Ling,~X.; Yu,~L.; Huang,~S.; Hsu,~A.~L.; Lee,~Y.-H.; Kong,~J.;
  Dresselhaus,~M.~S.; Palacios,~T. {Dielectric screening of excitons and trions
  in single-layer Mo{S}$_2$}. \emph{{Nano Letters}} \textbf{2014}, \emph{14},
  5569--5576\relax
\mciteBstWouldAddEndPuncttrue
\mciteSetBstMidEndSepPunct{\mcitedefaultmidpunct}
{\mcitedefaultendpunct}{\mcitedefaultseppunct}\relax
\EndOfBibitem
\bibitem[Guo \latin{et~al.}(2017)Guo, Sampat, Zhang, Robinson, Rupich, Chabal,
  Gartstein, and Malko]{Guo.2017b}
Guo,~T.; Sampat,~S.; Zhang,~K.; Robinson,~J.~A.; Rupich,~S.~M.; Chabal,~Y.~J.;
  Gartstein,~Y.~N.; Malko,~A.~V. {Order of magnitude enhancement of monolayer
  Mo{S}$_2$ photoluminescence due to near-field energy influx from nanocrystal
  films}. \emph{{Sci. Rep.}} \textbf{2017}, \emph{7}, 41967\relax
\mciteBstWouldAddEndPuncttrue
\mciteSetBstMidEndSepPunct{\mcitedefaultmidpunct}
{\mcitedefaultendpunct}{\mcitedefaultseppunct}\relax
\EndOfBibitem
\bibitem[Wang \latin{et~al.}(2018)Wang, Zhang, Liu, Li, Liu, Luo, and
  Ge]{Wang.2018}
Wang,~T.; Zhang,~Y.; Liu,~Y.; Li,~J.; Liu,~D.; Luo,~J.; Ge,~K.
  {Layer-Number-Dependent Exciton Recombination Behaviors of Mo{S}$_2$
  Determined by Fluorescence-Lifetime Imaging Microscopy}. \emph{{J. Phys.
  Chem. C}} \textbf{2018}, \emph{122}, 18651--18658\relax
\mciteBstWouldAddEndPuncttrue
\mciteSetBstMidEndSepPunct{\mcitedefaultmidpunct}
{\mcitedefaultendpunct}{\mcitedefaultseppunct}\relax
\EndOfBibitem
\bibitem[Robert \latin{et~al.}(2016)Robert, Lagarde, Cadiz, Wang, Lassagne,
  Amand, Balocchi, Renucci, Tongay, Urbaszek, and Marie]{Robert.2016}
Robert,~C.; Lagarde,~D.; Cadiz,~F.; Wang,~G.; Lassagne,~B.; Amand,~T.;
  Balocchi,~A.; Renucci,~P.; Tongay,~S.; Urbaszek,~B.; Marie,~X. {Exciton
  radiative lifetime in transition metal dichalcogenide monolayers}.
  \emph{{Phys. Rev. B}} \textbf{2016}, \emph{93}\relax
\mciteBstWouldAddEndPuncttrue
\mciteSetBstMidEndSepPunct{\mcitedefaultmidpunct}
{\mcitedefaultendpunct}{\mcitedefaultseppunct}\relax
\EndOfBibitem
\bibitem[Dou \latin{et~al.}(2020)Dou, Hu, Wang, Wang, Jin, Zhang, Shi, and
  Kou]{Dou.2020}
Dou,~K.~P.; Hu,~H.~H.; Wang,~X.; Wang,~X.; Jin,~H.; Zhang,~G.-P.; Shi,~X.-Q.;
  Kou,~L. {Asymmetrically flexoelectric gating effect of Janus transition-metal
  dichalcogenides and their sensor applications}. \emph{{Journal of Materials
  Chemistry C}} \textbf{2020}, \emph{8}, 11457--11467\relax
\mciteBstWouldAddEndPuncttrue
\mciteSetBstMidEndSepPunct{\mcitedefaultmidpunct}
{\mcitedefaultendpunct}{\mcitedefaultseppunct}\relax
\EndOfBibitem
\end{mcitethebibliography}

\end{document}